\documentstyle[12pt]{article}
\begin{document}
\font\bss=cmr12 scaled\magstep 0
\title{\vspace{2mm} Darboux-covariant differential-difference operators and dressing chains}

\author{\bss S. B. Leble ,\\
\small
Faculty of Applied Physics and Mathematics\\
\small Technical University of Gdan'sk,  \\
\small  ul. G.Narutowicza, 11/12 80-952, Gdan'sk-Wrzeszcz, Poland,\\
\small   email leble@mifgate.pg.gda.pl\\
\small and\\
\small Kaliningrad State University, Theoretical Physics Department,\\
\small Al.Nevsky st.,14, 236041 Kaliningrad, Russia.\\}
\vskip 1.0true cm

\date {}
\renewcommand{\abstractname}{\small Abstract}
\maketitle
\begin{abstract} The general approach to chain equations derivation for the function generated by
a Miura transformation analog is developing to account evolution (second Lax equation) and illustrated for
 Sturm-Liouville differential and
difference operators. Polynomial differential operators case is
investigated. Covariant sets of potentials are introduced by a
periodic chain closure. The symmetry of the
 system of equation with respect to permutations of the potentials
 is used for the direct construction of  solutions of the chain equations.
 A "time" evolution associated with some Lax pair is
 incorporated in the approach via closed t-chains. Both chains are combined
 in equations of a hydrodynamic type. The approach is next developed to general
Zakharov-Shabat differential and difference equations, the example
of 2x2 matrix case and NS equation is traced.

\end{abstract}

\section{ Introduction}

A potential reconstruction within some linear problem frame
depends on a class in which one searches a solution of inverse
problem \cite{Sab}. The coefficients (potentials) of the equation
itself naturally enter some algebraic structure generated by
transformations that preserve the form of the equation operator.
So the general variety of potentials splits to subsets invariant
with respect to the action of the transformations. Such
transformations as Darboux (Schlesinger, Moutard) ones (DT) are
generated by transformations of eigenfunctions $\phi$ of a given
differential (in D) operator. The DT introduces naturally some
intermediate object $ \sigma = (D\phi) \phi^{-1}$ that is linked
to a potential by the generalized Miura transformation. The main
ideas of G.Darboux, in principle, determine the form of the
transformation \cite{D} and the connection between potentials and
$\sigma$ via a factorization of the operator under consideration
\cite{FG,LeZa}. The structure of the transformation depends on the
ring to which the operator coefficients belong as well as abstract
differentiation realization \cite{L}, \cite{Mat}. The
transformations and the natural Miura transformation for
nonabelian entries (differential ring) are studied in \cite{LeZa}.
Namely the object allows to operate effectively by the spectral
problem data. We restrict ourselves here by one-dimensional
problems, but important steps towards the multidimensions are made
\cite{ABI,SabD}.

If one substitutes the potentials in terms of the $\sigma$ for any
iteration of the DT it yields in  chain equations that are
differential-difference equations and may be treated as the
equivalent form of the spectral problem one starts with. It was
discovered for the simplest one-dimensional problem  in \cite{We}
and extended in \cite{Sha}. The representation gives new
possibilities to produce explicit solutions as well as study
difficult questions of a potential uniform approximation
\cite{Novik}. The technique is directly connected with the quantum
inverse problem \cite{ChaSa} and its use in soliton equation
integration. In the next section we consider a general spectral
problem, polynomial in a differentiation. We begin from the
appropriate evolution equation and reduce the consideration to the
stationary case that generates a spectral problem. Showing how the
connection of the polynomial coefficients (potentials) with
$\sigma$ appears, we derive examples of the dressing chain
equation (DCE).

The solution of a periodically closed DCE may be analyzed from the
point of view of its bihamiltonian structure \cite{VSha} as, for
example, in \cite{Bl}. Some important algebraic structures
connected with the problem are studied in \cite{FSV}. The symmetry
of the system is naturally connected with the DT and generates a
finite group that we use to simplify the problem. In the Sec.3 we
introduce projecting operators for the irreducible subspaces of
the symmetry group and the corresponding variables. in the example
we consider the transition to the new variables give a chance to
express solutions of the DCEs in quadratures.

The same function $\sigma$ satisfies an additional equation that
appears if one studies the DT-covariance of the evolution equation
of a Lax pair. A "time" evolution associated with some Lax pair is
 incorporated in the approach via closed t-chains, see also the Sec. 3.
 Both chains are combined   in equations of a hydrodynamic type.

The chain equations for matrix spectral problems as
Zakharov-Shabat (ZS) one were considered in \cite{Sh3,Le}, see
also \cite{Novik}. In the sections 5 we continue to study the
general case and in 6 we treat an example of NS equation. The last
section is devoted to the specific features of a DSE derivation
for a difference ZS problem.

\section{The Miura maps and chain equations}
 Let us consider the differential operator
\begin{equation}\label{1}
  L  = \sum_{n=0}^N a_n D^n
\end{equation}
 on a differential ring A,  $\psi \in A$, with
 coefficients (potentials) $a_n \in A$  and an evolution equation
\begin{equation}\label{2}
  \psi_y = L\psi.
\end{equation}
 Here the operator D is a differentiation by some
 variable (or abstract one) and $\psi_y$ is the derivative with respect
 to another one (see \cite{Mat,L} for details and generalizations). We
 also would denote
 $D\psi =
 \psi'$. The transformation of the solutions of the
 equation is taken in the standard Darboux form
\begin{equation}\label{3}
  \psi[1] = D\psi - \sigma \psi,
\end{equation}
 where
\begin{equation}\label{4}
  \sigma = \phi' \phi^{-1}
\end{equation}
 with a linear independent invertible solution $\phi$ of (\ref{2}).

 The transformation of the coefficients of the
 resulting operator
 $$
 L[1] = \sum_{n=0}^N a_n[1] D^n
 $$
 is defined by
\begin{equation}\label{5}
  a_N[1] = a_N
\end{equation}
 and for all others n, by
\begin{equation}\label{6}
  a_n[1] = a_n + \sum_{k=n+1}^N [a_k B_{k,k-n} +
 (a_k' - \sigma a_k) B_{k-1,k-1-n}]
\end{equation}
that yields a covariance principle. It means that
 the function $\psi[1]$ is a solution of the
 equation
 $$
 \psi_y[1] = L[1]\psi[1].
 $$
 The result is a compact reformulation of Matveev theorem
 \cite{Mat}.

 The functions $B_{m,n}$ are introduced in \cite{LeZa}.
  We reproduce here the definition and useful relations for these differential polynomials.

 {\bf  Definition }
 $$
 B_{n,\,0}(\sigma) = 1,\quad n=0,1,2,\dots,
 $$
 {\it and recurrence relations}
 $$
 B_{n,\,k}(\sigma)=B_{n-1,\,k}(\sigma)+DB_{n-1,\,k-1}(\sigma),\quad
 k=\overline{1,\,n-1},\quad n=2,3,\dots.
 $$
 $$
 B_{n,\,n}(\sigma)=DB_{n-1,\,n-1}(\sigma)+B_n(\sigma),\quad
 n=1,2,\dots.
 $$
 {\it here and below the functions $B_{n}(\sigma)$ are standard Bell
 polynomials. It yields}
 $$B_{n,\,k}(\sigma)=\sum_{i=0}^{k-1}\,{n-i \choose
 n-k+1}\,B_{n,\,i}(\sigma)\, D^{k-i-1}\sigma,\quad
 k=\overline{1,\,n},\enspace n=0,1,2,\dots.
 $$
 The formula gives the convenient algorithm for the evaluation of the
generalized  Bell polynomials.

 One results in the important conclusion:

 {\bf {Statement 1}}.\enspace {\it{If a function}}\thinspace
 $\sigma$
 {\it {satisfies the equation}}
\begin{equation}\label{7}
 D_{y}\sigma = \sigma_y = Dr+[r,\,\sigma]
\end{equation}
 {\it where $r=\sum_{n=0}^{N}\,a_{n}\,B_{n}(\sigma)$;
 the operator \enspace$L_{\sigma} = D-\sigma$\enspace {\it{intertwines the operators}} %
 \enspace$D_y-L$\thinspace\ and $D_y- L[1]$.}

 Note that (\ref{7}) the for $\sigma$ from
 (\ref{4}) is the identity.

For the derivation of the dressing chain equations we consider the
stationary  solution of the evolution equation (\ref{2}) of the
theorem:
 \begin{equation}\label{mu}
  D_{y}\phi = \phi \mu,
\end{equation}
It gives
 $
 D_{y}\sigma=0.
 $
 For example, in matrix case $\mu = diag\{\mu_1,...,\mu_n\}$.
 Hence the corollary of (\ref{7})
\begin{equation}\label{8}
  Dr+[r,\,\sigma]=0,
\end{equation}
 is the Riccati equation analogue, mentioned in the introduction,  that we would name the generalized Miura map. It
 connects the ''potentials'' (coefficients of the operator L, see the expression for (\ref{1}))
 and $\sigma$ at
 every step (i) of DT iterations. Further we would supply the functions by the upper
  index i that show the number of iterations made. In the scalar case the commutator is zero and
 (\ref{8}) reads $r=c=const$. If one have the only potential $a_{0}$ with the rest
 invariant $a_{n}^{i}=a_{n}^{i+1}$, $n \neq 1$, then the expression for the
 potential is a direct generalization to one mentioned in the introduction:
\begin{equation}\label{9}
  a_{0}^{i}= - \sum_{n=1}^{N}a_{n}^{i}\,B_{n}(\sigma_{i})+c_{i}.
\end{equation}
 In this case the derivation of the chain equation is made by the substitution of (\ref{9})
 into the last of (\ref{6}) (n = 0), taken with the indices   i,i+1 at the
 correspondent side. The equation (\ref{9}) in the case N=2 gives the link of the
  $ \sigma_i$
 with the potential $u_i$ that enter the second order operator
\begin{equation}\label{10}
  L= -D^2 + u_i .
\end{equation}
 Supplying the relation by indices, one have
\begin{equation}\label{11}
   \sigma_{i,x}+\sigma_i^2+\mu_i = u_i.
\end{equation}
 Next nontrivial examples are connected with the case N=3 and the Miura transformation
 generalization (\ref{8}) is given by
\begin{equation}\label{12}
  \sigma_{i,xx}+\frac{3}{2}(\sigma_i^2)_x + \sigma_i^3+u_i\sigma_i+w_i = \mu_i.
\end{equation}
 It connects the coefficients $u_i, w_i$ with $\sigma_i$
 For both cases  the DT has the same form
 $$u_{i+1} = u_i - 2\sigma_{i,x}, $$
(for N=3 $a_3 = 1$, hence -2 should be changed to +3). Finally, if
one starts from the second order Sturm-Liuville equation
(\ref{10}), its  chain equation partner is
\begin{equation}\label{13}
\sigma_{i+1,x}+\sigma_{i+1}^2+\mu_{i+1} =
-\sigma_{i,x}+\sigma_i^2+\mu_i.
\end{equation}

For the third order operator  if $w_i=0$, the DCE is
\begin{equation}\label{14}
  (\sigma_{i+1,xx}+\frac{3}{2}(\sigma_{i+1}^2)_x + \sigma_{i+1}^3 - \mu_{i+1}) \sigma_i
 =(\sigma_{i,xx}+\frac{3}{2}(\sigma_i^2)_x + \sigma_i^3  - \mu_i)\sigma_{i+1}+
3\sigma_{ix}\sigma_i\sigma_{i+1}
\end{equation}
The case of zero $w_i$ is obviously a reduction for the space of
solutions of the linear problem and some modification of the DT
formula for the eigen functions is necessary \cite{LU}.

If a Lax pair of a nonlinear system is traced, one should consider one more operator, say,
 \begin{equation}\label{A}
  A = \sum_{n=0}^M b_n D^n
\end{equation}
and the correspondent evolution
 \begin{equation}\label{t-ev}
   \psi_t = A\psi.
\end{equation}
The coefficients of both equations (\ref{1},\ref{A}) depend on a
set of variables (potentials) $u_1,...u_n$ - eventual solutions of
nonlinear equations. In the case of joint covariance property
\cite{L,Le} the DT transforms of the coefficients induce the DT of
the potentials. The Statement 1 analogue accounts that the same
$\sigma$ enter the operator $L_{\sigma}$

{\it Statement 2. \enspace {\it{If a function}}\thinspace
 $\sigma$
 {\it {satisfies the equations}} (\ref{7}) and the equation}
 \begin{equation}\label{st}
 D_{t}\sigma = \sigma_t = Dq+[q,\,\sigma]
\end{equation}
 {\it where $q=\sum_{n=0}^{M}\,b_{n}\,B_{n}(\sigma)$;
  the operator \enspace$L_{\sigma} = D-\sigma$\enspace {\it{intertwines the pairs of operators}} %
 \enspace$D_y -L, D_t - A$\thinspace\ and $D_y - L[1], D_t- A[1]$.
 It means the integrability of the compatibility condition of the
equations (\ref{2},\ref{t-ev}) in the sense of the symmetry
existence with respect to DT of the potentials $u_i \rightarrow
u_i[1] $.}

 The compatibility condition of the equations (\ref{7}, \ref{st})
 yields the extra equation
\begin{equation}\label{cc}
  Dq_y+[q_y,\,\sigma]+ [q,\,Dr+[r,\,\sigma]] =
  Dr_t+[r_t,\,\sigma]+ [r,\,Dq+[q,\,\sigma]]
\end{equation}
That links the potentials and the element $\sigma$. In the case of
the unique potential it is possible to express the potential $u$
as a function of $\sigma$ \cite{LeB}. Considering the iterated
potentials
\begin{equation}\label{f}
  u_i = f_i(\sigma_i)
\end{equation}
 (now the index is again a number of iterations),
allows to produce the dressing chain equation substituting the
function into the DT formula $u_i[1] = u_{i+1} $.

The scalar case is much more simple. From (\ref{cc}) it follows
that $D(q_y -
  r_t) = 0$, or
\begin{equation}\label{ccs}
q_y - r_t = \sum_{n=0}^{M}\,(b_{ny}\,B_{n}(\sigma)+
b_{n}\,B_{n}'(\sigma)Dq) - \sum_{n=0}^{N}\,(a_{nt}\,B_{n}(\sigma)+
a_{n}\,B_{n}'(\sigma)Dr) =  const.
\end{equation}
The good example of this case is the KP equation and its dressing
chain, whence the potential is extracted from (\ref{ccs}).
 In the theory of solitons (integrable equations) the case (\ref{14})
generates Sawada-Kotera equation
 while (\ref{12}) corresponds to the famous KdV.
 Other reductions are more complicated
 from the point of view of the chain equations derivation: it is need to express the potential
 from (\ref{12}); e.g. so it is for the Boussinesq  equation reduction case
 \cite{LU}. For the chain equation in this case see \cite{LeB}.

\section{Periodic closure and time evolution}
The periodic closure of the DSE (\ref{13}) in the KdV case
produces a finite system of equations that possesses  the
bi-Hamiltonian structure \cite{VSha}. As the authors of the paper
had written about the case N=3 (one-gap potentials) "It is a
useful exercise to derive explicit formulas for  $\sigma_{i}$
directly from equations of the chain (\cite{We})." Below we
briefly show how to do it and give the formula. There is an
important question arises on this direct way: how to extract a
potential dependence on the additional parameter t  from a Lax
pair for a compatibility condition?

  We propose
to specify the "time" evolution via the t-dependence of
x-conserved quantities. In this section we begin to study this
problem in the terms of the same chain variables to be dependent
already on the time. Let us start from the system for three
functions $\sigma_{i}$ for the simplest nontrivial closure
\begin{equation}\label{15}
  \begin{array}{c}
    \frac{d}{dx}\sigma_{1}(x) =\ \sigma_{3}(x)^{2}-\ \sigma_{2}(x)^{2}+\mu_{3}-\mu_{2}, \\
    \frac{d}{dx}\sigma_{2}(x) =\ \sigma_{1}(x)^{2}-\ \sigma_{3}(x)^{2}+\mu_{1}-\mu_{3}, \\
    \frac{d}{dx}\sigma_{3}(x) =\ \sigma_{2}(x)^{2}-\ \sigma_{1}(x)^{2}+\mu_{2}-\mu_{1}. \
  \end{array}
\end{equation}
 The direct way in the bi-Hamiltonian formalism is initiated by the paper \cite{VSha}
 via generating function and conservation quantities introduction. In this simplest example we
 have taken now the integrals solve the problem completely. If one
 express the third variable $\sigma_{3}$ as the linear combination of
  the rest ones by means of the first integral (Casimir
 function) $c =  \sigma_{1}+\sigma_{2}+\sigma_{3}$ and substitute it into the
rest equations
 of (\ref{15}) then, after use of the second integral $ A$,
 one arrives to the differential equation
 of the first order for elliptic functions. It is convenient to show this fact in the terms
 of other variables \cite{VSha}
 $
 g_{1}(x)=\ \sigma_{1}(x)+\sigma_{2}(x),
 $
 $g_{2}(x)=\sigma_{2}(x)+\ \sigma_{3}(x),
$ $ g_{3}(x)=\sigma_{3}(x)+\ \sigma_{1}(x).$
 Let us exclude $g_{3}(x)$ by the connection
 $g_{3}(x)=2c-g_{1}(x)\ -g_{2}(x). $
 Further, omitting the argument x, the inverse transformations become
 $
 \sigma_{1} = -g_{2}   +c,
$
 $
 \sigma_{2} = - c + g_{1}  + g_{2},
 $
 $
 \sigma_{3} = c - g_{1}
 $
Plugging the transforms into the system (\ref{14}) we obtain two
differential  equations for the new variables
\begin{equation}\label{17}
  \frac{d}{dx}g_{1} =\allowbreak \mu _{1}-\mu _{2}+2cg_{1}   -g_{1}^{2}-
 2g_{1}g_{2},\\
 \frac{d}{dx}g_{2}
 =\ -2cg_{2} +2g_{1} g_{2} +g_{2}^{2}
 +\mu _{2}-\mu _{3}.\
\end{equation}
 The second integral of motion in terms of $g_i$ is more compact
\begin{equation}\label{19}
  A = g_{1} g_{2} g_{3} +\ \mu _{2}g_{3} +\allowbreak \mu _{1}g_{2} +\mu
  _{3}g_{1}.
\end{equation}
It allows to express $g_2$ as the function of $g_1$ and one
arrives at
\begin{equation}\label{20}
\begin{array}{c}
  \frac{d}{dx}g_{1}(x)=-\mu _{2}+\mu_{1}+2cg_{1} -g_{1}  ^{2}-
2(- g_{1}^{2}+\mu_{2}-\mu_{1}+2g_{1}c+ \\
   \left( g_{1}^{4}-4g_{1}^{3}c+
 2g_{1}^{2}(\mu _{1}+2c^{2}-2\mu_{3} +  \mu _{2})
 +4( A - \mu _{2}c- \mu_{1}c )g_{1}+ \mu_1^2+ \mu _{2}^{2}-2\mu_{2}\mu
 _{1}\right)^{1/2}.
\end{array}
\end{equation}

 The next problem is to solve the equation (\ref{20}) by means of the theory of elliptic functions.
 The Weierstrass or  Legendre canonical form of the integral
  yield a solution of the problem after Abel transformation \cite{Bat} and use of the algebraic
 formulas that give $\sigma_{2,3}$. Finally we have the explicit
 dependence of $\sigma_i$ on x and parameters c and A \cite{LeBr}, see also Sec.5.

 Let us turn to the problem of a time evolution arising from a Lax representation
 for some nonlinear equation. The main instruction is the search of the time dependence of
 the x-independent entries c and A of the solution of the equation (\ref{20}).
 Let us consider a DT-covariant evolution with the "time". In the KdV case the second Lax operator
 has the form (\ref{1}). The account of
 a connection between the potential and $\sigma$ (\ref{11})
produces the MKdV equation for the function $\sigma$. The
substitution into the equation the x-derivatives from the system
(\ref{15}) yields
\begin{equation}\label{21}
  \begin{array}{c}
    \frac{\partial }{\partial t}\sigma_{3}\left( x,t\right) =
 \frac{1}{2}\left( \mu _{1}+\mu _{2}-5\mu _{3}-F\right) \frac{%
\partial }{\partial x}\sigma_{3}(x,t),\\
F=\sigma_{2}^{2}+2\sigma_{2} \sigma_{3} +2\sigma_{1} \sigma_{2}
+\sigma_{1} ^{2}+2\sigma_{1} \sigma_{3} +\sigma_{3} ^{2}.
  \end{array}
\end{equation}
The equations for the $\sigma_{2.3}$ are written similar if one
use the mentioned cyclic symmetry (in indices). Such equations are
general; we would name them the t-chain equations. The form of the
equations is typical for the so-called hydrodynamic-type equations
. The system is diagonal and Hamiltonian, it could be integrated
by the hodograph method \cite{Ts}. The integration is trivial on a
subspace of c=const, for $F=c^2$. The final step of this direct
construction is the use of the equation (\ref{21}) and ones for
 $\sigma_{2.3}$. First of all one can check that
 $$
  \frac{\partial }{\partial t}c =
  -3\sigma_{2}^{2}\mu _{3}+3\sigma_{3}^{2}\mu _{2}-3\sigma_{3}^{2}\allowbreak \mu _{1}+
  3\sigma_{1}^{2}\mu _{3}-3\sigma_{1}^{2}\mu _{2}+3\sigma_{2}^{2}\allowbreak \mu _{1}
 =3c_x,
 $$
 that is zero if the x-dependence of the $\sigma_i$ is governed by (\ref{15}). If one  plugs
 the t-derivatives
 of $\sigma_{i}$ from formulas like (21) into the derivative $A_t$ so that
 $$
 A_t=
 =g_{2}^{2}g_{1}^{2}(3\mu _{1}- 3\mu _{3})+g_{2} ^{2} g_{3} ^{2}(3\mu _{2}-3\mu _{1})+
 g_{1} ^{2}g_{3} ^{2}(3\mu _{3}-3\mu _{2})+
 $$
  $$
 g_{2} g_{1} (6\mu _{3}\mu _{2}-6\mu _{1}\mu _{2})+g_{2} g_{3} (6\mu _{3}\mu _{1}-
 6\mu _{3}\mu _{2})+\allowbreak g_{1} g_{3} (6\mu _{1}\mu _{2}-6\mu _{3}\mu _{1})+
 3\left( \mu _{1}-\mu _{3}\right) \left( \mu _{2}-\mu _{3}\right) \left( \mu _{2}-\mu _{1}\right)
 $$
 The analysis of the expressions  leads to the statement

 {\bf Statement 2.} {\it If for all $g_i$ exists a common period $X$
the x-independent polynomial  c does not depend on t and A is linear function of t.}

For the proof it is enough to notice that the second order combination
$g_{2} g_{1} (6\mu _{3}\mu _{2}-6\mu _{1}\mu _{2})+g_{2} g_{3} (6\mu _{3}\mu _{1}-
6\mu _{3}\mu _{2})+\allowbreak g_{1} g_{3} (6\mu _{1}\mu _{2}-6\mu _{3}\mu _{1})$
is linear combination of x-derivative and constant, because, for example
$$ g_{1} \left( g_{2} -g_{3} \right) =
\frac{d}{dx}g_{2}(x)+\frac{d}{dx}g_{3}(x)+\mu _{2}-\allowbreak \mu _{1}
    $$
 After integration by x over the period X of the last
 equation  and use of the conservation
  laws
 for the KdV and MKdV equations in combination with (\ref{14}) one arrives to the linear dependence of
 $AX$
 on t. The coefficient may be recognized  in the expression for $A_t$: it is a combination of
 the eigenvalues
 $\mu_i$. For example, the $\int\sigma_{i}, \int\sigma_{i}^2, i = 1,2,3$ and other more
 complicated
 conserved  quantities \cite{Wad} should be accounted, the integrals over X of the combinations
are zero. The resulting formula for A should be substituted
 into the solution of (\ref{20}).

 {\bf Remark 1}. It is shown (see (\ref{21})) that the derivatives of $\sigma_k$ by t are
 linked with the derivatives by $x$ so that the equations of
 hydrodynamic type appear. The equations could be integrated by
 the methods developed in \cite{Ts}.

 \section{Discrete symmetry}
 \subsection{General remarks}
 One can easily check the invariance of the every conserved quantity (e,g, mentioned here c and A) against the
 permutations of the elements $\sigma_i$ as
 well as the covariance of the systems (\ref{15}) and (\ref{21}).
 The symmetry with respect to the cyclic permutation of the variables (or indices) is obvious
 if one recalls its DT origin, hence, the observation is general. For the illustration we again exploit the simplest "KdV" case
 starting
 from the representation of \cite{VSha,{FSV}}. Let us consider the periodically closed DCE of the odd N=2g+1 and
  introduce a vectorial notations, namely
 $$
 \Sigma = (\sigma_{1},...,\sigma_{N})^T, \qquad \mu =(\mu_{1},...,\mu_{N})^T
 $$
 and
 $$
  \Sigma^2 = (\sigma_{1}^2,...,\sigma_{N}^2)^T.
  $$
  The closed chain equation (\ref{13}) may be rewritten either in the form
$$
  (1+S)\Sigma_x = (1-S)(
   \Sigma^2  + \mu).
   $$
 or
\begin{equation}\label{23}
  \Sigma_x =  \sum_{k=1}^{N-1} (-1)^kS^k (\Sigma^2-\mu).
\end{equation}
 where the operator of permutation is represented as the matrix S \cite{VSha}. Both forms are
 obviously invariant with respect to the S-transformation because the matrix S and operators in the
 equations  (\ref{23}) commute.

 The same statement is valid for the equations of the time evolution.

 Let us emphasize that the r.h.s. of the equations (\ref{23}) is tensorial with respect to
 the components of $\Sigma$ so the action of the
 group transformation is the tensor (direct) product of the group representations in correspondent
 vector spaces.

 If one introduce the cyclic permutation operators  $T_s $, its
 action determine the matrix $S$ as
$$
 T_s\sigma_{i}=S_{ik}\sigma_{k}=\sigma_{i+1(modN)},
 $$
 the powers of the
 matrix of the previous section produce the group, $S^k \in C_n  \subset S_n$ such
 that is valid for any symmetry group and give a basis for integration of the covariant
 equations.
 The technique uses the Poisson representation of the
 system (\ref{23})
 \begin{equation}\label{Poi}
    \psi_x(\Sigma, \mu) = \{H(\Sigma, \mu),  \psi(\Sigma, \mu) \},
\end{equation}
where the operator
\begin{equation}\label{H}
    H(\Sigma,\mu) = \sum _{k=1}^{n}(\frac{1}{3}\sigma_i^3 + \mu_i\sigma_i)
\end{equation}
is invariant with respect to the group transformations and defines
a linear operator $ad_H$ with respect to the Poisson bracket
\begin{equation}\label{PB}
\{\sigma_{i},\sigma_{k}\} = (-1)^{k-i}(1-\delta_{ik}), k \leq i.,
\end{equation}
It is easy to check, that
$$\{H(\Sigma),   \sigma_j \} = \sum _{k=1}^{n}(\frac{1}{3}\{\sigma_k^3,\sigma_j \} + \mu_k\{\sigma_k,\sigma_j
\})=
 \sum _{k=1}^{n}( \sigma_k^2  + \mu_k)\{\sigma_k,\sigma_j \},
$$
that yields the system (\ref{23}).

The integration of the system may be understood in a "quantum
mechanical" language, introducing first the "commuting" functions
$C_i(\Sigma)$ from the operator $ad_H$ kernel $C_i \in K$:
\begin{equation}\label{K}
 ad_H C_i = 0.
\end{equation}
Next, the eigenvalue problem for a $\psi(\Sigma, \mu)$ outside the
kernel can be considered as a matrix one at some basis
\begin{equation}\label{nonK}
 ad_H  \psi_i =  \lambda_i \psi_i,
\end{equation}
$i=1,...g, n=2g+1$.
 The symmetry of
the equations (\ref{K}, \ref{nonK}) with respect to the
transformations $T_s$ (\ref{10}) follows from obvious relation
\begin{equation}\label{sym}
H(T_s\Sigma,T_s\mu) = H(\Sigma,\mu),
\end{equation}
it means that  matrices of a representation of the symmetry group
commute with the matrix of $ad_H$ on the correspondent subspace
  with constant Casimir operator.

  \subsection{Irreducible subspaces}

  The symmetry and the  tensor structure of the r.h.s. of the
  equation (\ref{23}) shows that the system may be simplified in the framework of
  Wigner-Eckart theory \cite{Wig}. The statement of the Wigner theorem reads in the context of the
  equations (\ref{K},\ref{nonK}) as the quasi-diagonal structure of the operator
  $ad_H$. Such structure and the correspondent simplification of
  the equations (\ref{K},\ref{nonK}) solution is achieved by a
  transition to the basis of irreducible representations.
The projecting operators $p_{i}$ to the irreducible subspaces are
defined in
  the subspaces produced
  by chains that appears as the sum over group of the transformation action to some basic
  element. In the case of commutative group, chosen for a
  simplicity, the irreducible matrices are one-dimensional, and
 the basis is defined by the set of projectors
\begin{equation}\label{pi}
  p_{i} = \sum_{s \in G} N_iD^i(s)T_s,
\end{equation}
  where $N_i$ are normalizing constants, $D^i(s)$ are irreducible representations of the symmetry
  group and the $T_s$ is the group transformation operator in the space to be considered.

  In our
  case   the operator coincide with the operator $T_s$ that have been introduced in the previous section
  and in the case of the cyclic permutation group $C_n$ the irreducible matrices are one-dimensional.
  Namely $D^j(e)=1, D^j(s)=a_j, D^j(s^2)=a_j^2, ...$
  with
  $a_j = exp(j 2\pi \imath/N)$ where the integer N is the group
  order. Hence
  \begin{equation}\label{IR}
    T_s s_j = a_js_j.
\end{equation}

  In the case of N=3 we project
  the system (23) or, originally (14) onto each subspace, having three equivalent
  equations.
Let it be $a=exp(2 \pi \imath/3)$; for example, the second
  of the resulting equations gives
\begin{equation}\label{26}
\begin{array}{c}
  s_{2x} = n_{1}+an_{2}+a^{2}n_{3} = \\
  \sigma _{3} ^{2}-   \sigma _{2} ^{2}+\mu _{3}-\allowbreak \mu _{2}+
   a\left( \sigma _{1} ^{2}-
\sigma _{3} ^{2}+\mu _{1}-
\mu _{3}\right)  \\
   +\allowbreak a^{2}\left( \sigma _{2} ^{2}-
   \sigma _{1} ^{2}+\mu _{2}-\mu _{1}\right).
\end{array}
\end{equation}
  Here $n_i$ denotes the r.h.s. of the equations (\ref{14}).

  The inverse transform  from original variables to the basis of irreducible
  representations reads
  $$\sigma _{1}  =s_{1}+s_{2}+s_{3},$$
  $$\sigma _{2}  =s_{1}+as_{2}+a^{2}s_{3},$$
 $$\sigma _{3}  =s_{1}+a^{2}s_{2}+as_{3}.$$
Plugging into the equation (\ref{26}) yields
$$\frac{d}{dx}s_{1}(x)=0,$$
\begin{equation}\label{ss}
\begin{array}{cc}
\frac{d}{dx}s_{2}(x)= -\left( a-1\right) \left( \ \allowbreak
3a\left( s_{2}^{2}+
 2s_{1}s_{3}\right) -a\mu _{2}+a\mu _{1}-\mu _{2}+\mu _{3}\right),  \\
 \frac{d}{dx}s_{3}(x)=\allowbreak \left( a-1\right) \left( \ 3a\left( s_{3}^{2}+
2s_{1}s_{2}\right) -a\mu _{3}+a\mu _{1}+\mu _{2}-\mu _{3}\right).
\end{array}
\end{equation}

The second conservation law (\ref{19}) in terms of $s_i$ allows to
express the hamiltonian as a function of the only variable ($s_3
$).  The conservation laws are obviously the combinations of the
irreducible polynomials $\sigma_i$.Together with the Hamiltonian
(=$\lambda$) conservation it leads to the spectral curve
definition.

Returning to general problem  needs the tensor product space of
the vectors $\Sigma, \mu$. The problem of solution of the
equations (\ref{K},\ref{nonK}) simplifies if one use the mentioned
symmetry,
 written in terms of
\begin{equation}\label{si}
    s_i =\sum_{s \in G} N_iD^i(s)T_s\sigma_1= N^{-1}\sum_{k=1}^N a_{i-1}^{k-1}
    \sigma_k.
\end{equation}

{\bf Statement 3} {\it By a direct application of the operator
$T_s$ it could be checked that the tensor products of $s_i$ (see
(\ref{IR}))
\begin{equation}\label{tp}
  s_is_k...s_j,
\end{equation}
form a basis of irreducible tensors in the space of polynomials:
the result differs from (\ref{tp}) by a constant factor $ a_i
a_k...a_j$.}

The further computations are conveniently made via the Poisson
bracket
\begin{equation}\label{PB}
\{s_{j},s_{l}\} = N^{-2} \sum_{ik}a_{j-1}^{i-1}a_{l-1}^{k-1}
\{\sigma_{i},\sigma_{k}\}.
\end{equation}

Particularly it is easy to show  that the $C_i$ could be chosen as
a combination of irreducible polynomials (one could check this
statement at the level of the examples from the mentioned paper
\cite{VSha}), the presented conservation laws are  combinations of
the irreducible polynomials of $\sigma_i$.

Some $ad_H$-invariant chains are constructed by the following
algorithm. We established that the first irreducible combination
is proportional to the Casimir operator
\begin{equation}\label{s1}
  ad_H  s_1 = 0,
\end{equation}
Let us return again to the simplest example of N=3, taking the
second equation of the basic system (e.g. (\ref{si}))
\begin{equation}\label{s2}
 ad_H  s_2 =  n_2 = \left( a-1\right) \left( \ 3a\left( s_{3}^{2}+
2s_{1}s_{2}\right) -a\mu _{3}+a\mu _{1}+\mu _{2}-\mu _{3}\right)
\equiv n^{(1)},
\end{equation}
with the polynomial expression in the r.h.s..  Acting next, as
\begin{equation}\label{n2}
 ad_H  n_2 =  n_2'\equiv n^{(2)},
\end{equation}
one produces the higher order polynomial ar r.h.s; repeating,
\begin{equation}\label{nr}
 ad_H  n^{(i)} =    n^{(i+1)},
\end{equation}
one arrive at the statement which express
\begin{equation}\label{psi}
    \psi = \sum_0^{\infty} n^{(i)}, n^{(0)} = s_1, n^{(1)}=s_2
\end{equation}
as a polynomial function of $s_2$ and $\lambda$.

{\bf Statement 4 \label{ps}} {\it The eigenfunction $\psi$ of the
operator $ad_H$ is expressed as the function of the variable
$\lambda$ via solution of the algebraic (quadratic) equation}
\begin{equation}\label{ae}
 \lambda \psi = \psi - s_2 - s_1.
\end{equation}
The proof is delivered by the construction, taking into account
the link (\ref{19}) by which we express the powers $s_2^k$ with
$k\leq 3$.

An interesting link to the theory of automorphic functions
applications to the system (\ref{23}) could be found via the same
symmetry. The example of (\ref{15}) transformed as (\ref{ss})
allows to demonstrate it in almost obvious way. Let us recall that
it is necessary to consider x as a complex variable. If one
applies the operators $T_s$ or $T_s^2$ to the both sides of the
first and second equation of the system (\ref{ss}) and,
simultaneously substitute
\begin{equation}\label{Meb}
 x = a^{-1}x',
 or x = a^{-2}x',
\end{equation}
 the equation rest invariant. In addition to the symmetry the
 double-periodicity of the solutions of the equations as elliptic
 functions (see also (\ref{nonK}) or its Jacobi version  by the transition $\psi_i =
 \exp[\zeta_i]$) yields in $s_i(x')=s_i(x)$,
 \begin{equation}\label{auto}
    x' = \alpha x + \tau; \alpha =1,a,a^2; \tau=n_1\tau_{1}+n_2\tau_2.
\end{equation}
The property have place for any genus g, $\alpha \in C_{2g+1}$.
The solutions could be built in Poincare $\theta$-series
\cite{Bern}. See also \cite{Bor}.

 Next, the t-dependence may be introduced via the
scheme of Sec.3. Other nonlinear systems are treated similar.
Widen symmetry, for example from \cite{FSV} includes reflections
at the $\Sigma,\mu$ tensor product space and could give more
information about solutions.

 \section{Oprerator Zakharov-Shabat problem}

Let us reformulate the general scheme of the dressing chain
derivation in the nonabelian case from \cite{Le}, starting from
the evolution
\begin{equation}\label{51}
  u \Psi + J D\Psi = \Psi _{t},
\end{equation}
with the polynomial first order L(D)-operator. The case
 is the nontrivial  example of a general equation (\ref{2}) with operator entries
  J and u (y is changed to t ).
 This way a form of evolution operator
(Hamiltonian) is fixed as $J D + u$. Such form allows to consider
one-dimensional Dirac equation
 \cite{Yu}, or,
 in a stationary case, one goes down to a multilevel
system interacting with a quantum field \cite{MR1689163}. We would
respect $\Psi$ or other (necessary for DT construction) solution $
\Phi$ as operators. The equation could enter to the Lax pair of
some integrable nonlinear equations as Nonlinear Schr\"odinger or
Manakov ones.

 The potential $u$ may be
 expressed in the terms of $\sigma$ from  the equation
 (\ref{7}), that we would rewrite as
\begin{equation}\label{Miu}
- \sigma_t+ J \sigma_x + [J \sigma, \sigma] = [\sigma, u]
\end{equation}
The structure of this equation determine the algebraic properties
of the admissible dressing construction.

 The x-stationary version, when $u_x, \sigma_x = 0,$, the equation
 \ref{Miu} yields
\begin{equation}\label{sMiu}
- \sigma_t + [J \sigma, \sigma] = [\sigma, u]
\end{equation}
that means $(Sp \sigma)_t = 0$ and the traceless  possibility. The
structure of $\sigma$ imply also the restriction
\begin{equation}\label{DetM}
Det\sigma = Det M = \prod\mu_i
\end{equation}
  Namely, introducing the iteration index i, we have the link

\begin{equation}\label{52}
  Du^{i}+[u^{i},\sigma_{i}] = -DJ-[J, \sigma_{i}]\sigma_{i},
\end{equation}
 the connection is linear, but contains the commutator. Let us denote
$ad_{\sigma}x=[\sigma,x]$. Then
\begin{equation}\label{53}
  u^{i}=(ad_{\sigma_{i}} - D)^{-1}\left( DJ + [J,\sigma_{i}]\sigma_{i}\right).
\end{equation}
 The existence of the inverse operator in (\ref{53}) need some restriction for the
 expression in () brackets, the expression should not belong to the kernel
 of the operator $D-ad_{\sigma_{i}}$. In the subspace, where the Lie product is zero,
 the equation (\ref{52}) simplifies.
 The DT is also simple.
\begin{equation}\label{54}
  u^{i+1}=u^{i}+[J,\sigma_{i}]
\end{equation}
 Note that $J$
 is not changed under DT (see (\ref{5})
 Substituting the link (\ref{53}) for i,i+1 into (\ref{54}) one arrives to the chain
 equations. One also could express matrix elements of $u$ in terms of the
 elements of the matrix $\sigma$ and plug it into the Darboux transform (\ref{54})
 separately.

 Let us give more details of the construction in the stationary case, restricting $DJ=0$
 and $\Psi, \Phi$ correspond to  $\lambda, \mu$.
 Note that there are two possibilities for stationary equations from nonabelian (\ref{51}):
  either
  $$
  \Psi _{t} = \lambda \Psi
  $$
  or
\begin{equation}\label{55}
  \Psi _{t} = \Psi \lambda
\end{equation}
  and the first of them  leads to the essentially   trivial connection between solutions
  and potentials from the point of view of DT theory \cite{Mat}.

In the second case one writes
\begin{equation}\label{56}
  \sigma = \Phi_x\Phi^{-1}=J^{-1}(\Phi\mu\Phi^{-1}-u)
\end{equation}
and DT takes the following form in terms of $s_i$
\begin{equation}\label{57}
  u^{i+1}J^{-1}u^{i}J - J^{-1}[s^i,J],
\end{equation}
where it is denoted $s = \Phi\mu\Phi^{-1}$, here and further
iteration number indices omitted. The potentials $u^i$ may be
excluded from the equation (\ref{7}) for this case
$$
\sigma_t=Dr+[r,\sigma],
$$
with
\begin{equation}\label{58}
  r=J\sigma+u=s.
\end{equation}
The stationary case, after the plugging $u^i$ from (\ref{58}) and
returning indices, gives
\begin{equation}\label{59}
  s^{i+1}=s^{i}+J \sigma^{i+1} -\sigma^{i}J
\end{equation}
and
\begin{equation}\label{510}
  Ds^{i}+[s^{i},\sigma^{i}] = 0.
\end{equation}
links the derivative of  s and the internal derivative of
$\sigma$. The formal transformation that leads to the chain
equations is similar to the (\ref{53}), namely, after substitution
$$
\sigma^{i}=-ad_{s_i}^{-1}Ds^i
$$
 into the equation (\ref{59}).

 Further progress in the development of this programm is connected with the choice
 of the additional algebraic structure over the field we consider. It can be useful for
 the concrete representation of solutions of the equation (\ref{510}).
 For example, if the elements
 $s^{i},\sigma^{i}$ belong to a Lie algebra with structure constants
 $C_{\alpha \beta}^\gamma$, then, after the choice of the basis $e_\alpha$ one
 introduces the expansions (summation over the Greek indices is implied))
 $$
 s^{i}=\xi_\alpha^i e_\alpha
 $$
 and
 $$
  \sigma^{i}=\eta_\alpha^i e_\alpha.
 $$
 Plugging into (\ref{510}) gives the differential equation
 $$
 D\xi_\alpha^i+C_{\gamma \beta}^\alpha\xi_\gamma ^i\eta_\beta^i=0.
 $$
 If one defines the matrix
\begin{equation}\label{511}
B_{\beta\alpha} = C_{\gamma \beta}^\alpha\xi_\gamma ^i,
\end{equation}
 then, outside of the kernel of B
 \begin{equation}\label{eta}
  \eta_\beta^i=-B^{-1}_{\beta\alpha}D\xi_\alpha^i.
\end{equation}
 By the definition of the Cartan subalgebra C the correspondent subspace does not
 contribute in the Lie product of (\ref{510}).

 {\bf Statement.} {\it
 If, further, J belongs to a module over
 the Lie algebra, $Je_\alpha = J_{\beta\alpha}e_\beta$ and there exist an
external involutive
 automorphism $tau$ such that determine $e_{\alpha}J$ (e.g. $(ab)^{\tau} = b^{\tau}a^{\tau})$. Then,
 the chain equation for the variables $\xi_\alpha^i $
 takes the form}
 $$
 \xi_\alpha^{i+1} =
 \xi_\alpha^i  - B^{-1}_{\beta\gamma}(D\xi_\gamma^{i+1}) J_{\beta\alpha}+
 (J_{\alpha\beta}^{\tau}B^{-1}_{\beta\gamma}D\xi_\gamma^i)^{\tau},
 $$
 {\it where the matrix B is defined by (\ref{511}) and the components
$e_\alpha$ outside of C}.
 {\it Otherwise
 $$
 D\xi_\gamma^i=0,
$$
 if $e_\gamma \in C$}.
 {\bf Proof} The statement is in fact the DT in the form of the
 equation (\ref{59}) written in the basis of $e_{\alpha}$, in
 which the expression (\ref{eta}) is used. The subspace of C gives
 the second case.

The system of differential equations
 is hence nonlinear as the matrix B
 depends on $\xi_\gamma ^i.$

 {\bf Remark.} The scheme may be generalized for  a nonstationary equation (2.5).
 The equation (5.2)
 then have the additional term $D\sigma^{i}_t$ from the r.h.s.

\section{\bf Example of NS case }

The split NS equation is produced by the Lax pair based on
(\ref{51}) of the second order
\begin{equation}\label{ZS2}
\Psi_x = i
\left(%
\begin{array}{cc}
  \lambda & p \\
  q & -\lambda \\
\end{array}%
\right)\Psi,
\end{equation}
while the NS equations is solved under the  reduction $p =
\bar{q}$. The choice $J = \sigma_3$ is obviously used.

 Let
$$
\sigma = \eta_i\sigma_i, u = u_1\sigma_1+u_2\sigma_2 \in sl(2,C)
$$
with the Pauli matrices $\sigma_i,$ as the basis $e_{\alpha}$
\begin{equation}\label{alg}
  [\sigma_i,\sigma_k] = 2\imath\varepsilon_{iks}\sigma_s,
\end{equation}
 as generators of $sl(2,C)$. The matrix realization determine the
 left and right actions of J as usual matrix product, so the
 condition of the Statement of the previous section is satisfied.
The "Miura" connection (\ref{sMiu}) is specified by
\begin{equation}\label{sJ}
   J\sigma = i\sigma_2\eta_1 - i \sigma_1\eta_2 + \eta_3\sigma_0.
\end{equation}
Plugging the result and $u, \sigma$ into the (\ref{sMiu}), one
arrives at the equations:
\begin{equation}\label{etas}
    \begin{array}{c}
      \eta_1' + 2\eta_1\eta_3 = 2\imath \eta_3 u_2, \\
        \eta_2' + 2\eta_2\eta_3 = - 2\imath \eta_3 u_1,  \\
         \eta_3' - 2\eta_1^2 - 2\eta_2^2 = 2\imath \eta_2 u_1 - 2\imath \eta_1 u_2. \\
    \end{array}
\end{equation}
From \ref{DetM} it follows
\begin{equation}\label{eta3}
 \eta_3 = \sqrt{\mu_1\mu_2 - \eta_1^2 - \eta_2^2},
\end{equation}
that is in accordance with (\ref{etas}). The chain equations hence
are
\begin{equation}\label{ch}
  \begin{array}{c}
    u_1^{i+1} = u_1^{i} - 2i \eta_2^i \\
  u_2^{i+1} = u_2^i + 2i\eta_1^i \\
  \end{array}
\end{equation}
where
\begin{equation}\label{ui}
  \begin{array}{c}
    u_1  =  - (\eta_2'/\eta_3 + 2\eta_2)/2i \\
     u_2 = (\eta_1'/\eta_3 + 2\eta_1)/2i  \\
  \end{array}
\end{equation}
 the correspondent iteration index $i$ is implied and $\eta_3$ is
  the function (\ref{eta3}).
 The structure of the matrix elements of the potential $u$
is such that $q=u_1+iu_2$ so the reduction to NS case means the
reality of both $u_i$. The repulsive NS corresponds to $p = -q^*$.
The similar scheme is developed in \cite{Sh3}.

\subsection{\bf The chain closures}
Simplest version  of the closure, at the "first level' gives
\begin{equation}\label{clo}
    \eta _s^{i+1} = \alpha_s\eta^{i}_s.
\end{equation}
Introducing notations,
$$
\eta_1^{1} = x, \eta_2^{1} = y, \eta _3^{1} = z,
$$
and taking into account the conditions (\ref{DetM}) yield
\begin{equation}\label{xyz}
\begin{array}{c}
  -\mu^{1}_1\mu^{1}_2= x^2+ y^2+z^2, \\
   -\mu^{2}_1\mu^{2}_2=\alpha_1^2x^2+\alpha_2^2y^2+\alpha_3^2z^2,\\
\end{array}
\end{equation}
providing restrictions for the constants
$$
-\mu^{1}_1\mu^{1}_2= x^2+ y^2-(\mu^{2}_1\mu^{2}_2-\alpha_1^2x^2-\alpha_2^2y^2)/\alpha_3^2,
$$
or
$$\alpha_3^2= \alpha_1^2 = \alpha_2^2, \mu^{1}_1\mu^{1}_2\alpha_3^2 = \mu^{2}_1\mu^{2}_2$$,
hence
\begin{equation}\label{alf}
\alpha_1=\pm\alpha_3,\alpha_2=\pm\alpha_3.
\end{equation}
The equations (\ref{ch}) go to
\begin{equation}\label{eq}
\begin{array}{c}
    (\frac{\alpha_1}{\alpha_3} - 1)x_t/z = 2(\alpha_1+1)y,\\
(\frac{\alpha_1}{\alpha_3} - 1)x_t/z = 2(\alpha_1+1)x
\end{array}
\end{equation}
excluding z, one arrives at
 $$\frac{\alpha_2 - \alpha_3}{1 + \alpha_2}ln_t(y) =
 \frac{\alpha_3 - \alpha_1}{ 1 + \alpha_2}ln_t(x),$$
 in a non-trivial conditions for constants (\ref{alf}),
$$
\alpha_1=-\alpha_3,\alpha_2=-\alpha_3,\\
 ln_t(y) = - ln_t(x),
$$
and, finally, gets
\begin{equation}\label{y}
y = c/x.
\end{equation}
Plugging (\ref{y}) into (\ref{eq}) and then $z$ into (\ref{xyz})
yields in
\begin{equation}\label{alf1}
\frac{x_t}{\sqrt{-\mu^{1}_1\mu^{1}_2x^2 - x^4  - c^2}} =
-(1-\alpha_3)
\end{equation}
that is again solved in elliptic functions \cite{Bat}. The
t-chains are obtained in a way similar to the section 3 using the
second Miura map from Sec 2.

\section{Nonlocal operators}

 Let again A be an operator ring, with the automorphism T.
If for any two elements $f,g \in A$
$$
T(fg)= T(f)T(g),
$$
general formulas for DT for polynomial in $T$ operators exist
\cite{Mat}, see also \cite{Lpr} for examples. Here we continue to
study the versions of ZS problem. We name the operator T as a
shift operator, but it could be general as defined above.

 Let us take the general evolution
equation in the case N=1.
\begin{equation}
\psi_{t}\left( x,t\right) =  \left( J+U T\right)\psi. \label{ZS}
\end{equation}
There are two types of DT in this case \cite{Lpr}, denoted by
indices $\pm$. The DT of the first kind (+) leaves $J$ unchanged.
We rewrite the transform of $U$ as
\begin{equation}
U^{+}=\sigma ^{+}\left( TU \right) \left( T\sigma ^{+}\right)
^{-1} \label{DTZ}
\end{equation}
where $\sigma^+ = \phi(T\phi)^{-1}$, further the superscript $"+"$
is omitted.

For the spectral problem correspondent to (\ref{ZS}), the
nontrivial transformations appear if in the stationary equation
 one introduces the constant element $\mu$ that does
not commute with $\varphi$ and $\sigma$:
\begin{equation}
 \left( J+U T\right) \varphi =\varphi \mu. \label{ZSP2}
 \end{equation}
 The formula for the potential is then changed as
\begin{equation} \ U =\varphi \mu (T\varphi)^{-1} - J\sigma.
   \label{S+}
\end{equation}
Let us derive the identity that links the potential U and
$\sigma$, doing it in a different manner than in \cite{Lpr} or,
here, in the Sec. 6, starting from
\begin{equation}\label{Ts}
  T(\sigma)T^2(\varphi) = T(\varphi),
\end{equation}
and plugging it into the shifted equation (\ref{ZSP2}):
\begin{equation}\label{TSZ}
   T(U) T^2(\varphi) = T(\varphi)\mu - JT(\varphi).
\end{equation}
One has a Miura-like link
\begin{equation}\label{MT}
    \sigma T(U) \sigma = U + [J,\sigma].
\end{equation}
$ T(\sigma) = \sigma^{-1}$ is accounted. Comparing with
(\ref{DTZ})   yields new form of DT, that coincides with
(\ref{DTn}).
\begin{equation}\label{DTn}
  U + [J,\sigma] = U^+.
\end{equation}
Direct use of the equation (\ref{ZSP2}) for expressing U in terms
of $\tau = \varphi\mu\varphi^{-1}$ and $\sigma$ gives
\begin{equation}\label{U}
 U = \tau\sigma - J\sigma.
\end{equation}
The element $\tau$ is useful, also, for
\begin{equation}\label{T(U)}
 T(U) = \sigma^{-1}\tau - J\sigma^{-1}.
\end{equation}
Plugging (\ref{T(U)}),(\ref{U}) into (\ref{MT}), one arrives at
identity.
 The algorithm of the explicit derivation of the chain equations
begins from the equation (\ref{MT}) solving with respect to U in
appropriate way. For matrix rings, it may be a system of equations
for matrix elements, that could be effective
 in low matrix dimensions of the "Miura" (\ref{MT}).

 The role of $\sigma^+$ can play also the function $s = \varphi \mu
(T\varphi)^{-1}$. The equation (\ref{S+}) connects U and $\sigma$.
Let us rewrite (\ref{MT}) and the DT  in terms of $s$, excluding U
from (\ref{S+}) , denoting the number of iterations by index
$$
U[n] = s_n -J\sigma_n.
$$
The equation  (\ref{MT}) reduces to
\begin{equation}
 s_n = \sigma_n T(s_n) \sigma_n \label{S}
 \end{equation}
 The use of this result gives for the DT
\begin{equation}
 s_{n+1} - s_n = J \sigma_{n+1} + \sigma_nJ.
\label{Ss}
 \end{equation}
 Then, solving the result (\ref{S}) with respect to $s$  one have the chain system.
 It could be made similar to the previous section by means Lie
 algebra representation.

Let us mention that the chain equations for the classical ZS
problem and two types of DT transformation were introduced in
\cite{SH}. The closure of the chain equations specify classes of
solutions.

\section{ Conclusion}

Concluding, we wish to express a feeling that the technique
elements we develop are general. Chain equation derivation is
simply the result of substitution of a potential as the function
of $\sigma$ into the DT formulas, but this problem of explicit
form of the function could be non-trivial. The periodic closures
of a chain for arbitrary N for KdV and other equations are studied
very similar and leads to the expressions for the $\sigma_i$ and,
consequently, for the potentials in hyper-elliptic functions by a
construction. We also believe that the finite closures for the
equations may produce the solutions by the similar combination of
a symmetry analysis for both x- and t- evolutions. The development
of the technique for infinite chains do not look impossible as
well.

 The work is supported by the Polish Ministry of
Scientific Research and information Technology grant
PBZ-Min-008/P03/2003. Author would like to thank M. Pavlov for
fruitful discussion.

\end{document}